\documentclass[aps, 10pt, prl, twocolumn, amssymb, amsmath, amsfonts, showpacs, superscriptaddress, preprintnumbers, a4paper]{revtex4-2} 
\usepackage{amsmath, amssymb, amsthm}
\usepackage{physics}
\usepackage{times}
\usepackage{graphicx, xcolor}
\usepackage[latin1]{inputenc}
\usepackage{comment}  
\usepackage[
  breaklinks,
  colorlinks=true,
  bookmarks=true,
  bookmarksopenlevel=3,
  bookmarksnumbered=true,
  menucolor=blue,
  linkcolor=blue,
  citecolor=blue,
  urlcolor=blue,
  pdfcreator={GhostScript},
  pdfkeywords={},
  pdfstartview=FitH]{hyperref}

\begin{document}

\def\lptms{Universit\'e Paris-Saclay, CNRS, LPTMS, 91405, Orsay, France.}
\def\lps{Universit\'e Paris-Saclay, CNRS, Laboratoire de Physiques des Solides, 91405, Orsay, France.}

\title{Kinetic formation of trimers in a spinless fermionic chain}

\author{Lorenzo Gotta}\email{lorenzo.gotta@universite-paris-saclay.fr}\affiliation{\lptms}
\author{Leonardo Mazza}\affiliation{\lptms}
\author{Pascal Simon}\affiliation{\lps}
\author{Guillaume Roux}\affiliation{\lptms}

\date{\today}

\begin{abstract}
We show the stabilization of two trimer phases in a chain of spinless fermions with a correlated hopping term.
A trimer fluid forms due to a gain in trimer kinetic energy and competes with a fluid of unbound fermions.
Furthermore, we observe two intermediate phases where these two fluids coexist and do not spatially separate.
Depending on the way trimers are created out of the Fermi sea, hybridization can occur, in which case the onset of correlations between the two fluids is well captured by a generalized BCS ansatz.
These results are finally extended to the formation of larger multimers, which  highlights the peculiarities of pair and trimer formation.
\end{abstract}

\maketitle

Pairing of more than two particles is a transverse topic in physics,
from nuclear and particle physics to condensed matter and cold-atom gases~\cite{Efimov1, Efimov2, Schuck2016, Guan_2013}.
Beyond neutrons and protons made of three bound quarks, trimers have been widely discussed in the context of Efimov states~\cite{EFIMOV1970, EFIMOV1973, Petrov2003,Piatecki_2014,Carlson_2017} and in Helium physics~\cite{kunitski2015observation, kolganova2017}.
With the versatility of cold-atom platforms in terms of internal degrees of freedom and interactions, many proposals for trimers formation arose in the quantum matter literature, using spin balanced~\cite{Akos_2007, Capponi_2008, Lecheminant_2008, Guan_2009, Azaria_2009, Backes_2012, Niemann_2012, Pohlmann_2013, Kornilovitch_2013} or spin-imbalanced fermionic mixtures~\cite{Kuhn_2012, Peng_2014, Szirmai_2017}, or fermions with different masses~\cite{Burovski_2009, Orso_2010, Roux_2011}. Signatures of bosonic trimers have also been discussed both in one~\cite{Keilmann_2009,Dalmonte_2012,Zhang_2014,Fan_2017} and two dimensions~\cite{You_2012,Zhang_2015,Guijarro_2020}.

Forming trimers composed of a unique fermionic specie -- spin-polarized fermions -- is particularly challenging despite seminal results in the context of the quantum Hall effect~\cite{Read_1999}.
The pairing of spinless fermions already shows a rich phenomenology~\cite{Moore_1991, Read_2000, Kitaev_2001, Mattioli_2013, Dalmonte_2015, Ruhman_2017, Kane_2017, He_2019, Gotta_2020, Gotta_2021, Gotta_2021_B}, and it is a crucial mechanism for some topological phases of matter, motivating further investigations in this direction.
An intuitive route is to use attractive density interactions~\cite{He_2019} on a chain and stabilize trimers using a third neighbor repulsion to prevent phase separation.
In order to develop a low-energy description, as a trimer phase cannot be interpreted as an instability of the Luttinger liquid theory, the authors of Ref.~\cite{He_2019}  propose an emergent-mode description, which is then treated with bosonisation tools. 
Unfortunately, such approach is not conclusive on the nature of the transition from the Luttinger liquid to the trimer phase.

In this Letter, instead of using interactions, we propose and study a simple microscopic model that realizes a trimer phase thanks to a trimer-hopping term, extending the pairing mechanism foreseen in Ref.~\cite{Ruhman_2017}.
It contains tightly bound trimers and allows for a detailed study of the transitions between the fermionic and the trimer phases.
Remarkably, we do not find direct transitions but two intervening coexistence phases that we describe using a two-fluid approach. Our analysis generalizes the pioneering work of Ref.~\cite{Kane_2017}, where pairing physics is described with the help of two fictitious fluids of unpaired and paired fermions, a method that was successfully applied~\cite{Gotta_2021, Gotta_2021_B} to the model of Ref.~\cite{Ruhman_2017}. The model offers an exciting playground in which the trimer density and effective interaction with the unbound fermions is controlled by a single parameter.

\begin{figure}[t]
\centering
\includegraphics[width=\columnwidth]{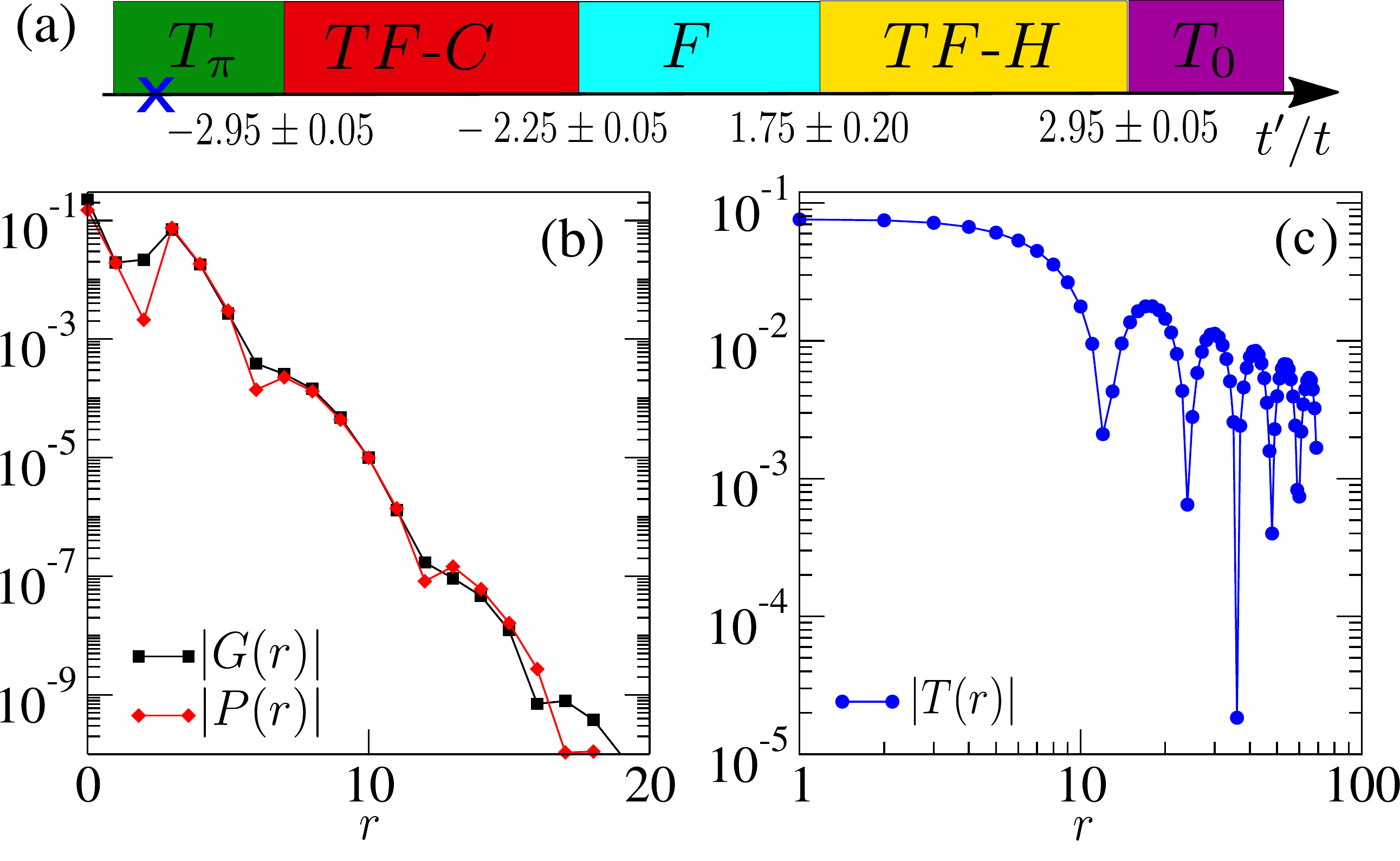}
 \caption{(a) Phase diagram of the model in Eq.~\eqref{Eq:Ham:GRA} (not in scale). (b,c) Correlators for $t^{\prime}/t=-3.5$ (blue cross in (a)): $G(r)$ for single-particle, $P(r)$ for pairs (both exponential) and $T(r)$ for trimers (algebraic). }
 \label{Fig:GRA:PhDi}
\end{figure}

\textit{Model and phase diagram --}
We consider a chain of size $L$ with $N$ spinless fermions described by creation and annihilation operators $\hat c_j^{(\dagger)}$.
We generalize the pair-hopping model of Ruhman and Altman~\cite{Ruhman_2017} by introducing a trimer-hopping term $t'$ competing with particle-hopping $t>0$:
\begin{equation}
 \hat H =  -t\sum_j \big( \hat c_j^\dagger \hat c_{j+1}  + \text{h.c.}
 \big)
 - t'\sum_j \big( \hat T_j^\dagger \hat T_{j+1}  + \text{h.c.} \big).
 \label{Eq:Ham:GRA}
\end{equation}
Here, $\hat T_j = \hat c_j \hat c_{j+1} \hat c_{j+2}$ is the trimer operator.
In what follows, we choose a fermion density $n = N/L = 0.25$ (we fix the lattice spacing to $1$).
As made of three fermions, trimers are fermions. In fact, the $\hat T_j$ operators satisfy anticommutation relations at lattice distances larger than two, as well as $\hat T_j^2 = 0$.
We compute the ground-state properties of~\eqref{Eq:Ham:GRA} using state-of-the-art density-matrix renormalization group (DMRG) simulations~\cite{White1992, White1993, Schollwock_2005, Schollwock_2011} implemented from the ITensor library~\cite{itensor}.

In order to systematically probe the phase diagram summarized in Fig.~\ref{Fig:GRA:PhDi}(a), we compute local observables and the single-particle, pair and trimer two-point correlators
\begin{align*}
 G(r) =  \langle \hat c_j^\dagger \hat c_{j+r}\rangle,\;
  P(r) = \langle \hat P_j^\dagger \hat P_{j+r}\rangle,\;
 T(r) = \langle \hat T_j^\dagger \hat T_{j+r} \rangle,
\end{align*}
with $\hat P_j =\hat c_j \hat c_{j+1}$.  When $t'/t = -3.5$,
Fig.~\ref{Fig:GRA:PhDi}(b-c) shows clear evidences for a trimer phase : the trimer correlator $T(r)$ displays an algebraic decay, whereas single-particle and pair correlators $G(r)$ and $P(r)$ decay exponentially. Moreover, the single-particle and pair correlation lengths coincide. This reflects the existence of both single and two-particle gaps, while the three-particle excitations remain gapless. The $t'>0$ and $t'<0$ phases have a different nature and are respectively denoted by $T_0$ and $T_\pi$ as we will see.
The $t'=0$ free fermions point extends in a regular fermionic Luttinger-liquid phase $F$ in which all correlators are algebraically decaying.
The analysis of the energy density and its derivatives as a function of $t'/t$ allows to find the boundaries of these five different phases: the two coexistence phases appear for $-2.95\pm 0.05< t'/t < -2.25\pm 0.05$ and $1.75\pm 0.20< t'/t < 2.95\pm 0.05$.

We first clarify the nature of the $T_{0,\pi}$ phases by studying the model for $t=0$, which is exactly solvable. We assume that the ground state only explores the trimer subspace $\mathcal H_T$ spanned by Fock states with clusters of $3n$ fermions with integer $n$. For instance, $\ket{\circ \bullet \bullet \bullet \circ \bullet\bullet\bullet\bullet\bullet\bullet}$ belongs to $\mathcal H_T$, whereas $\ket{\bullet\bullet\bullet\bullet \circ \circ \circ \bullet \circ \circ \bullet \bullet \bullet}$ does not. In order to describe $\mathcal H_T$, we introduce an effective lattice whose length includes an excluded volume associated to each trimer: for $N_T = N/3$ trimers, we consider $L-2 N_T$ sites. 
Within the subspace $\mathcal H_T$, the Hamiltonian reduces to a free fermion model on the aforesaid effective lattice and the ground state is a Fermi sea of trimers occupying the lowest-energy states of the dispersion relation $\varepsilon_T(k) = - 2 t' \cos(k)$. For $t'>0$, its minimum lies at $k_0=0$, whereas for $t'<0$ it lies at $k_0 = \pi$. The Fermi points are located at $k_0 \pm \pi  \frac{N_T}{L-2N_T}$.
The quantitative comparison of this effective model with exact numerical simulations is excellent, see Supplementary Material~\cite{SuppMat}. Depending on the value of $k_0$, we call $T_0$ and $T_\pi$ the two trimer phases. Moreover, this value plays a crucial role in the description of the transition to the $F$ phase, as we are now going to discuss.

\textit{The transition from $T_\pi$ to $F$ --}
In order to describe the rest of the phase diagram, we employ an effective many-fluid approach. 
We start with the study of the transition from the $T_\pi$ side ($t'<0$) to the $F$ phase by introducing a purely phenomenological two-fluid model. One fluid carries unbound fermions while the other carries the trimers. These two fluids are defined on two effective chains of length $L$ and  
are described by the two Hamiltonians
\begin{equation}
 \hat H_F = \sum_k \varepsilon_{F}(k)  \hat a_k^\dagger \hat a_k, \qquad
 \hat H_T = \sum_k \varepsilon_{T}(k) \hat d_k^\dagger \hat d_k,
 \label{Eq:TwoFluids}
\end{equation}
where $\varepsilon_{F}(k) = -2t \cos k$ and $\varepsilon_{T}(k)$ has been defined previously.
Note that $\hat a_k$ and $\hat d_k$ are fermionic operators and do not have any exact connection with the lattice operators $\hat c_j$.

\begin{figure}[t]
\includegraphics[width=\columnwidth]{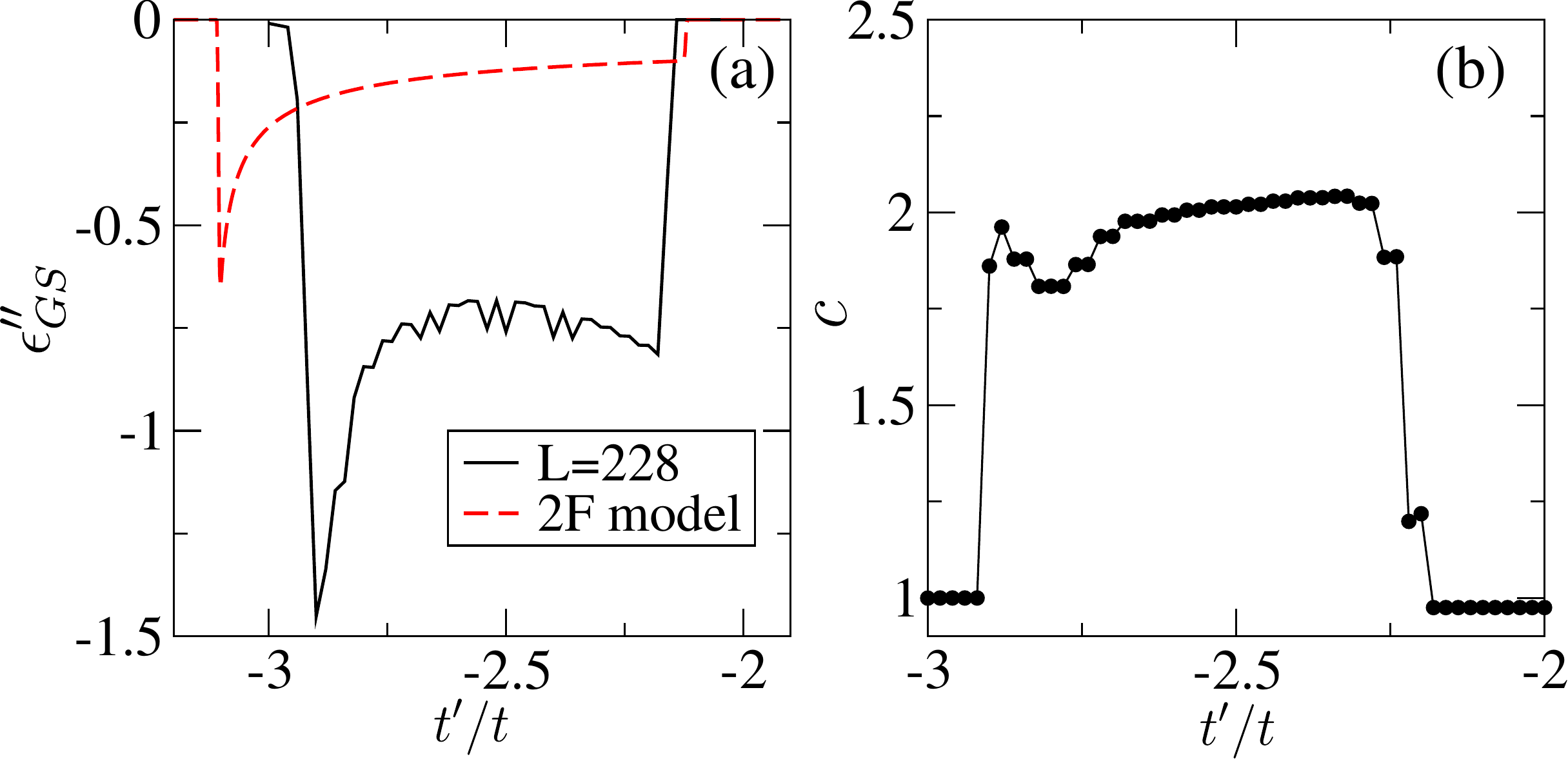}
 \caption{(a) Second derivative of the ground state energy density $\epsilon_{GS}^{\prime\prime}$
 for the Hamiltonian in Eq.~\eqref{Eq:Ham:GRA} (continuous line) and for the model $\hat{H}_{2F}$ (dashed line). (b) Central charge $c$ as a function of $t^{\prime}/t$.}
 \label{Fig:TFC}
\end{figure}

The phase diagram of $\hat{H}_{2F} = \hat H_F+\hat H_T $ captures all the qualitative aspects of the transition from $F$ to $T_\pi$. 
We obtain it by imposing the filling constraint $n = n_F + 3n_T$ and finding the optimal value for the trimer density $n_T$ that minimizes the total energy, $n_F$ being the unbound fermions density.
For $t'<0$, we find that the $T_\pi$ phase, such that $n_T=n/3$ and $n_F=0$, appears for $t^{\prime}/t < -3/\cos\left(\frac{\pi n}{3}\right)\simeq -3.11$. The $F$ phase, such that $n_T=0$ and $n_F=n$, appears for $t^{\prime} / t > -3\cos(\pi n) \simeq -2.12$. 
Between these two critical values, an intermediate phase emerges, in which both bands are partially populated. 
This \textit{trimer-fermion coexistence phase} ($TF$-$C$) is characterized by the presence of trimers and unpaired fermions that delocalize over the same spatial region in a miscible state.
A similar behavior was recently observed for pairs and unbound fermions in Refs.~\cite{Gotta_2021, Gotta_2021_B}.
The interpretation of the intermediate phase as a $TF$-$C$ is supported by the numerical data presented in Fig.~\ref{Fig:TFC}(a), where we compare the behavior of the second derivative of the energy density computed using $\hat H_{2F}$ with the one obtained from ground-state DMRG simulations for the Hamiltonian~\eqref{Eq:Ham:GRA}. 
In addition, the numerical calculation of the central $c$ charge~\cite{SuppMat} gives $c=2$ for $TF$-$C$ and $c=1$ for $T_\pi$ and $F$. This probes the number of gapless modes of the system and further supports the proposed two-fluid interpretation.
We thus conclude that two Lifshitz transitions characterized by a non-perturbative reshaping of the Fermi points separate the three phases for $t'<0$.

The remarkable agreement shows that the residual interaction between unbound fermions and trimers is negligible, so that the two fluids hardly hydridize. The reason is that the fermionic and trimer Fermi seas are significantly displaced in momentum space. Any interaction process responsible for turning a trimer into three fermions (or \textit{viceversa}) must conserve the lattice momentum, which makes the process extremely unlikely. At most, 
density-density interactions between fermions and trimers might only shift the locations of the phase boundaries.

\begin{figure}[t]
\includegraphics[width=\columnwidth]{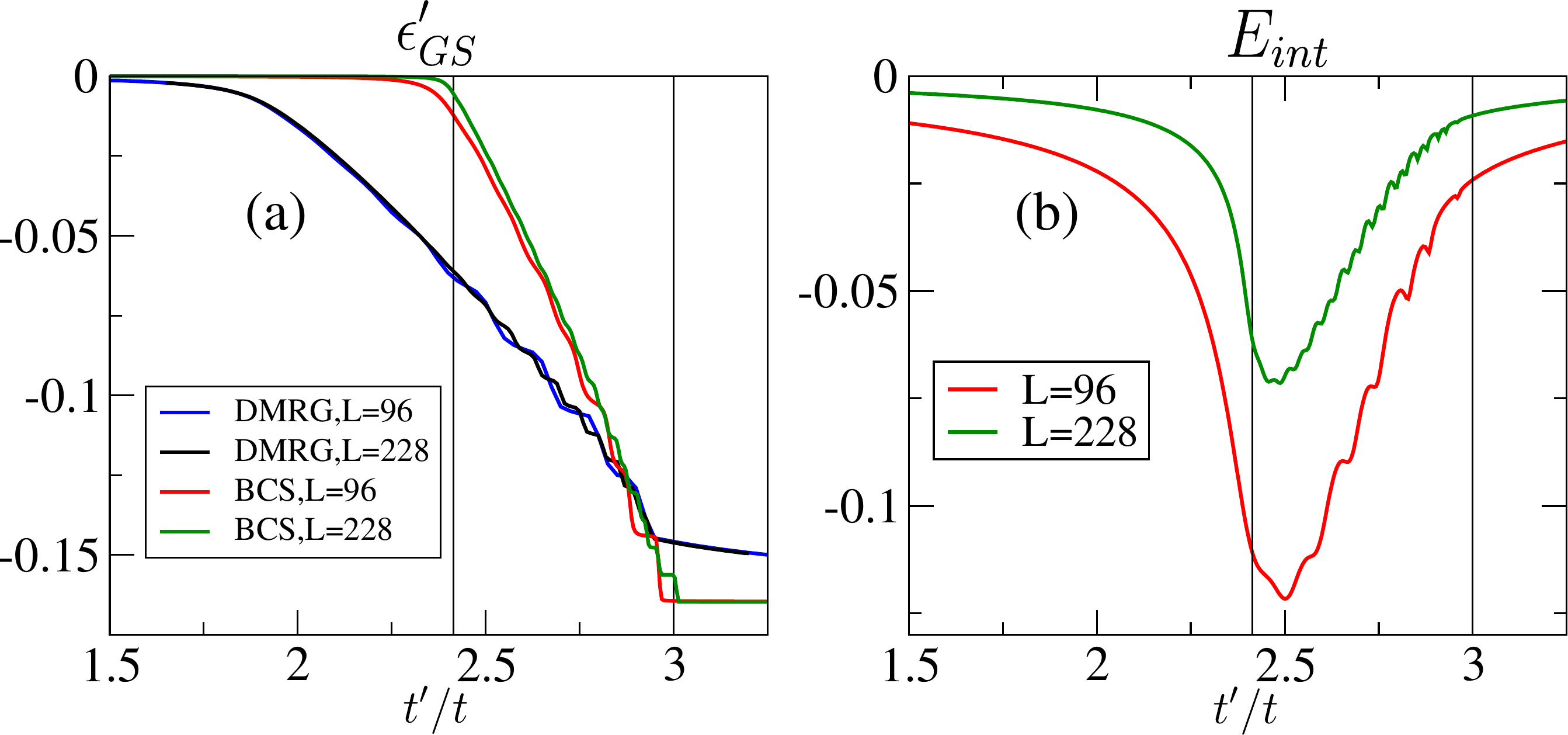}
\caption{(a) First derivative of the ground-state energy density $\epsilon_{GS}^{\prime}$ from DMRG and the 3BCS ansatz with $g/t=8$. (b) Interaction energy $E_{\text{int}}$ for the 3BCS ansatz using $g/t=8$.
Vertical lines are transition points predicted by the 3BCS model.} 
 \label{Fig:BCS:Study}
\end{figure}

\textit{A BCS-like approach for the transition from $T_0$ to $F$ --} Since both bands minima are at $k=0$ for $t'>0$, the study of the transition from the $F$ phase to the $T_0$ phase is more challenging. In this case, the two Fermi seas are overlapping and momentum-conserving processes turning fermions into trimers (and \textit{vice versa}) can take place. We need to include in the Hamiltonian a new term such that $\hat H_{2F} =  \hat H_F+ \hat H_T + \hat H_{\text{int}}$. To model the interaction, we consider a recombining term $\hat H_{\text{int}} =  g \sum_j d_j^\dagger \hat a_{j-1} \hat a_j \hat a_{j+1} + \text{h.c.}$.
The value of the parameter $g$ is introduced on a purely phenomenological basis.  

We propose an approach based on a BCS-like ansatz wavefunction, that we dub the 3BCS ansatz, which aims at capturing the emergence of the trimers.
First, we introduce $\ket{n_F}$ as the Fermi sea state with fermionic density $n_F=n$ and Fermi momentum $k_F = \pi n$. Concerning trimers, we start from the vacuum $\ket{v_T}$.
Our ansatz wavefunction reads:
\begin{equation}
 \ket{\Psi_3} = \hspace{-0.4cm} \prod_{-\frac{k_F}3<k<\frac{k_F}3} \hspace{-0.5cm}
 \left( 
 \alpha_k + \beta_k 
 \hat d_k^\dagger 
 \hat a_{-k_F+\delta_k} 
 \hat a_k 
 \hat a_{k_F-\delta_k}
 \right) 
 \ket{n_F} \hspace{-0.07cm} \otimes \hspace{-0.07cm} \ket{v_T},
 \label{Eq:BCS:3}
\end{equation}
where $\delta_k = 2k$ for $k\geq0$ and $\delta_k =-2k-2 \pi/L$ for $k<0$.
The wavefunction interpolates from the fermionic limit $\beta_k=0$ to the trimer phase limit $\alpha_k=0$. 
In the former case, the state reduces to $\ket{n_F}\otimes \ket{v_T}$, while in the latter case, trimers form a Fermi sea with Fermi momentum $k_T=\pi n/3$ and the fermionic system is empty.
When both $\alpha_k$ and $\beta_k$ are different from zero, $\ket{\Psi_3}$ includes quantum correlations between fermions and trimers: the creation of a trimer with momentum $k$ is accompanied by the annihilation of three fermions at momenta $k$ and $\sim \pm(k_F-2|k|)$. 
This choice is motivated by a minimization calculation  showing that in order to create a trimer with momentum $k=0$, the most favorable choice is to annihilate three fermions at $k=0$ and $k=\pm k_F$ (see~\cite{SuppMat} for a detailed discussion).
The ansatz~\eqref{Eq:BCS:3} thus captures some physically-relevant forms of quantum correlations between the two fluids, and satisfies momentum conservation in the trimer-fermion exchange process. In addition, $\ket{\Psi_3}$ offers the advantage of being entirely composed of commuting terms (like the standard BCS wavefunction) and is thus  well suited for analytical calculations. 

By virtue of the normalisation condition $|\alpha_k|^2+|\beta_k|^2=1$, we introduce the parametrisation $\alpha_k= \cos \theta_k$ and $\beta_k = e^{i \varphi_k} \sin \theta_k$, with $\theta_k \in [0,\pi]$ and $\varphi_k \in [0,2\pi[$. 
We compute the expectation value of the Hamiltonian $\hat H_{2F}$ in state $\ket{\Psi_3}$. Defining $\epsilon_{GS}=\frac{\langle \hat H_{2F} \rangle - E_{FS}}{Lt}$, where $E_{FS}$ is the energy of the fermionic Fermi sea filled up to momentum $k_F$, we obtain:
\begin{equation}
\epsilon_{GS} = 2
 \int_0^{\frac{k_F}{3}}\!\! \big[ 
 A_k \sin^2 \theta_k - B_k \sin 2 \theta_k \sin \varphi_k
 \big] \frac{\dd k}{2 \pi},
 \label{Eq:Energy:Functional}
\end{equation}
where: 
\begin{subequations}
 \begin{align*}
  A_k =& 2 (1- t'/t) \cos k + 4 \cos (k_F-2k), \\
  B_k =& \frac {2g}{Lt}\left[\sin(k_F-k)+\sin(4k-2k_F)+\sin(k_F-3k)\right].
 \end{align*}
\end{subequations}
Minimizing the functional~\eqref{Eq:Energy:Functional} yields the solutions
\begin{equation}
\theta_k = \frac 12 \arctan
 \left( 
 \frac{2B_k}{A_k}
 \right) + \frac \pi 2 
 \Theta \left(- A_k \right),\quad
\varphi_k = \frac{\pi}{2},
 \label{Eq:Optimal_Params}
 \end{equation}
where $\Theta(x)$ is the Heaviside step function.
The plot of the first derivative of the ground-state energy density obtained with the 3BCS ansatz is shown in Fig.~\ref{Fig:BCS:Study}(a), where we compare it with DMRG simulations of the Hamiltonian~\eqref{Eq:Ham:GRA}. 
By taking $g = 8 t$, we reproduce correctly the numerical results obtained with $L=96$ and $L=228$, apart from a quantitative mismatch in the first transition point.

The 3BCS ansatz reproduces two important features of the numerical $\epsilon_{GS}^{\prime}$ curve: first, the step-like behavior appearing close to the $T_0$ phase and, second, the smoother profile close to the F phase. The first is a finite size effect that originates from the absence of a strong fermion-trimer hybridization: each step corresponds to the formation of a trimer. 
The second is naturally associated to the fermion-trimer recombination processes, which are here strongly relevant.
This is further supported by the behavior of the fermion-trimer interaction energy $E_{\text{int}}= -\frac{L}{\pi}\int_0^{\frac{k_F}{3}} \! B_k \sin 2 \theta_k \dd k$ plotted in Fig.~\ref{Fig:BCS:Study}(b). We observe that
the step-like behaviour coincides with the weakly-interacting region, whereas the smoother one is associated to strong interactions.
Strictly speaking, the 3BCS ansatz gives an interaction energy $E_{\text{int}}$ that scales to zero in the thermodynamical limit~\cite{SuppMat}, as seen in Fig.~\ref{Fig:BCS:Study}(b). Yet, we believe that both in the DMRG and in the exact solution of $H_{\text{2F}}$, a residual hybridization remains relevant in the thermodynamical limit.
The above considerations motivate the wording \textit{trimer-fermion hybrid phase} ($TF$-$H$) for a phase that displays relevant hybridization processes at low trimer density on finite size systems and likely in the thermodynamical limit.

\begin{figure}[t]
\includegraphics[width=\columnwidth]{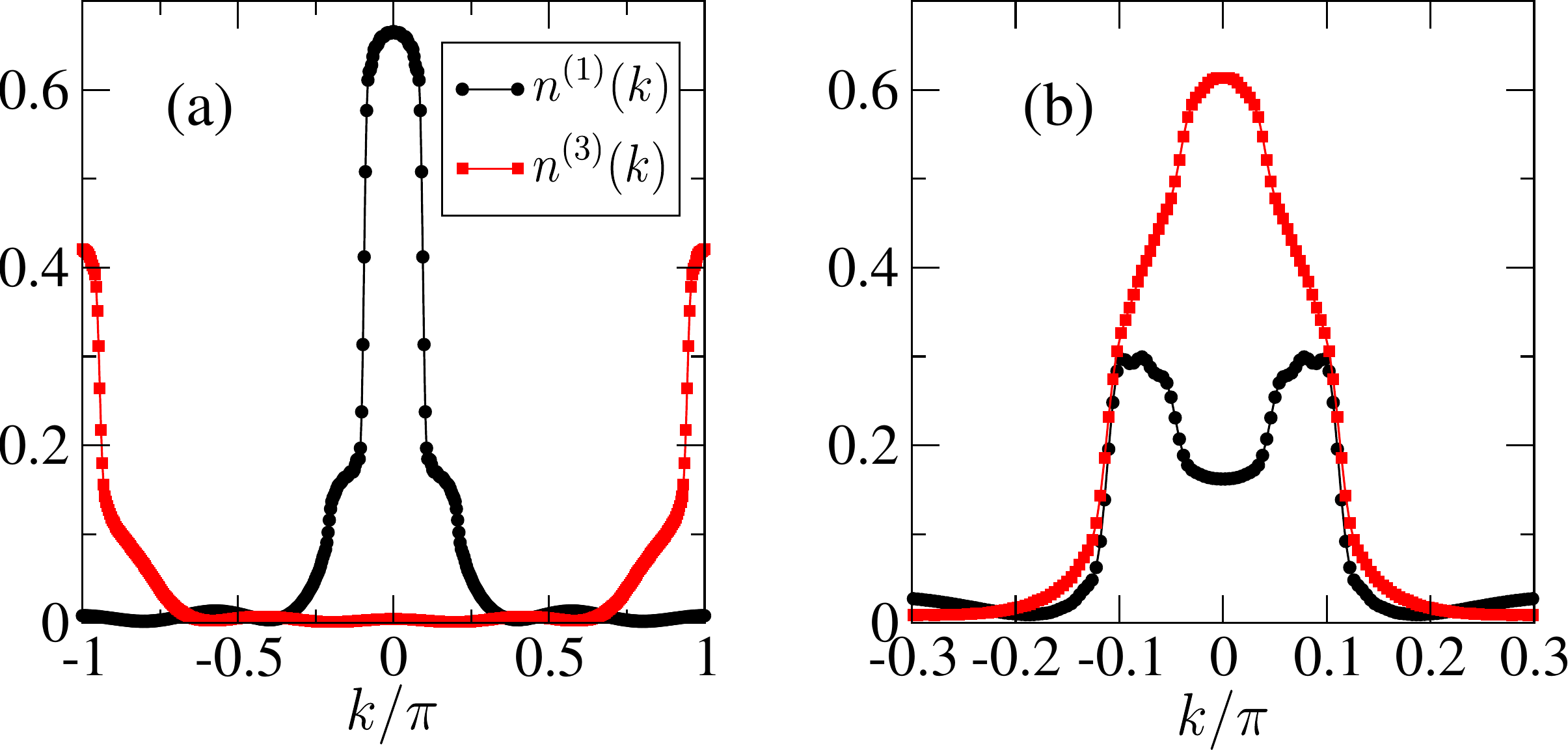}
 \caption{Momentum distribution functions $n^{(1)}(k)$ and $n^{(3)}(k)$ with OBC and $L=96$ in (a) the $TF$-$C$ phase for $t^{\prime}/t=-2.7$, and in (b) the $TF$-$H$ phase for $t^{\prime}/t=2.9$ restricted to $-0.3<k<0.3$.}
 \label{Fig:nk}
\end{figure}

To further support the relevance of the hybridization processes and their description with the 3BCS ansatz, we introduce "isolated" operators
\begin{equation}
 \hat F^{(M)}_j = (1-\hat n_{j}) \left( \prod_{m=1}^M \hat c_{j+m} \right) (1-\hat n_{j+M+1}).
 \label{Eq:Projected_Operators}
\end{equation}
that capture unbound fermions ($M=1$) and isolated trimers ($M=3$).
In Fig.~\ref{Fig:nk}(a-b), we present the DMRG momentum occupation functions $n^{(M)}(k) = \sum_{j,l} e^{- i k (j-l)} \langle \hat F^{(M)\dagger}_j \hat F^{(M)}_l\rangle$ computed from their two-point correlators. Their behaviors fully agree with the two-band picutre and highlight the main differences between the $TF$-$C$ and $TF$-$H$ phases. 
In the latter case, the unbound fermion distribution in black displays a remarkable hollow around $k=0$. Such a feature is perfectly coherent with the structure of the 3BCS ansatz, according to which the filling of the trimer states around $k=0$ occurs via the annihilation of fermions around $k=0$ and $k=\pm k_F$. While the fermionic occupation function is expected to develop a minimum at $k=0$, its trimer counterpart gets maximal at $k=0$. This feature is more visible close to the $T_0$ phase, where the transition is sharp and the trimer density higher, and less visible close to the $F$ phase.

\textit{Perspectives on tetramer formation --}
We conclude by conjecturing the ground-state phase diagram of the generalization of the  Hamiltonian~\eqref{Eq:Ham:GRA} to the case of multimer correlated hopping. We start with the case of tetramers:
\begin{equation}
 \hat H =  \sum_j \left( -t \hat c_j^\dagger \hat c_{j+1} +t' \hat M_j^\dagger \hat M_{j+1}
 + H.c.
 \right),
 \label{Eq:Ham:Tetramers}
\end{equation}
where $\hat M_j=\hat c_j \hat c_{j+1} \hat c_{j+2}\hat c_{j+3}$ is the tetramer operator. Models featuring $\hat M_j$ as the interaction term have been discussed, e.g., in the form of generalized Kitaev chains with even multiplet pairing fields~\cite{Mazza_2018}, leading to the prediction of non-topological parafermions.  
As in the trimer case, we find a standard $F$ phase when $|t^{\prime}|\ll t$ and two tetramer $M$ phases with quasi-long-range-ordered tetramer correlator $M(r)=\langle \hat M_j^\dagger \hat M_{j+r} \rangle$ and exponentially decaying $G(r)$, $P(r)$ and $T(r)$. 

\begin{figure}[t]
\includegraphics[width=\columnwidth]{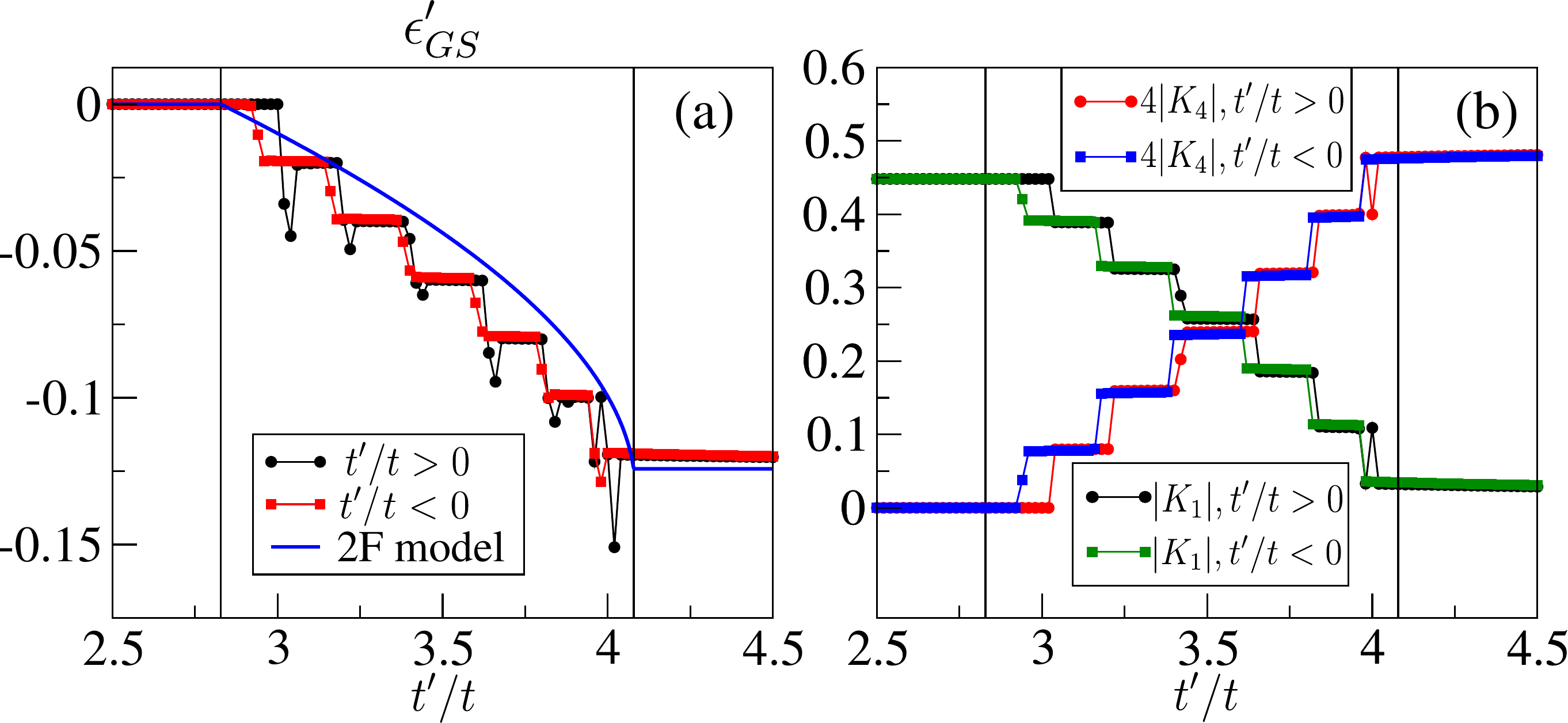}
 \caption{Energy observables in the ground-state of \eqref{Eq:Ham:Tetramers} as a function of $\abs{t^{\prime}/{t}}$ for both $t'\geq 0$ and $t^{\prime}\leq 0$.
Vertical lines are transition points predicted by the generalized two-fluid model.
(a) First derivative of the energy density from DMRG and two-fluid model. (b) Single-particle and tetramer kinetic energy densities $|K_1|$ and $4|K_4|$.}
 \label{Fig:tetramers_symmetry}
\end{figure}

In order to discuss the nature of the transition from the $M$ phases to the $F$ phase, we present in Fig.~\ref{Fig:tetramers_symmetry}(a-b) the first derivative of the ground state energy density, the total single-particle kinetic energy density $K_1=\frac{1}{L}\sum_j \langle \hat{c}^{\dag}_j \hat{c}_{j+1}+H.c.\rangle$ and the total tetramer kinetic energy density $K_4=\frac{1}{L}\sum_j \langle \hat{M}^{\dag}_j \hat{M}_{j+1}+H.c.\rangle$. Superimposing the curves obtained for $t^{\prime}/t<0$ and $t^{\prime}/t>0$ shows a remarkable quantitative agreement.
We conclude that the two transitions have the same properties, differently from what has been found for pairs and trimers~\cite{Ruhman_2017,Gotta_2021,Gotta_2021_B,SuppMat}.

The intervening phases between $F$ and $M_{0,\pi}$ are quantitatively described by a direct generalization of the non-interacting two-fluid model predicting a coexistence of unbound fermions and tetramers~\cite{SuppMat}.
The comparison with the DMRG data of Fig.~\ref{Fig:tetramers_symmetry}(a) shows an excellent agreement. Accordingly, two Lifschitz transitions separate the coexistence phase from the fermionic phase and the tetramer phase.
The numerical data indicate that the effective interactions between the two fluids are strongly suppressed, even when the tetramers quasi-condense at $k \sim 0$ for $t^{\prime}<0$.
A first qualitative argument is that the larger the molecule, the higher is the order in perturbation theory to split it into $M$ unbound fermions.
Aonther is that an heuristic argument supporting an emergent  $t^{\prime}\rightarrow -t^{\prime}$ symmetry in the Hamiltonian. 
Indeed, by considering an equivalent version of Hamiltonian~\ref{Eq:Ham:Tetramers} $\hat H_M(t,t^{\prime})=-t\sum_j \hat c^{\dag}_j\hat c_{j+1}-t^{\prime}\sum_j \hat c^{\dag}_j\left(\prod_{m=1}^{M-1}\hat n_{j+m}\right)\hat c_{j+M}+H.c.$ for a general $M$-particle correlated hopping term, we see that 
the unitary transformation $\hat c_j\rightarrow e^{i\frac{\pi}{M}j}\hat c_j$ transforms $\hat H_M(t,t^{\prime})$ into $\hat H_M(e^{i\frac{\pi}{M}}t,-t^{\prime})$. In the limit of large molecules $M\rightarrow +\infty$, the phase factor multiplying $t$ tends to $1$, connecting $\hat H_M(t,t^{\prime})$ to $\hat H_M(t,-t^{\prime})$. Since the coexistence phase is the generic scenario at low density when molecules quasi-codense at $k\sim \pi$, the same is expected for quasi-condensation at $k\sim 0$ on the opposite side.

\textit{Conclusions --}
In this Letter, we have studied the ground state phase diagram of a one-dimensional lattice model of spinless fermions with trimer hopping by means of an effective two-fluid model.
Two remarkable coexistence phases are found, with one in which hybridization between unbound fermion and trimers is well described by an effective BCS ansatz. 
This scenario is generalized and becomes generic for a class of Hamiltonian with an hopping of arbitrary large molecules.

In the context of bound states formation with spinless fermions,
the success of our interacting two-fluid approach  provides a new set of interpretative ideas and technical tools that are expected to shed further light on the properties of experimentally relevant Hamiltonians for which multimer formation has been predicted~\cite{He_2019}. The possibility of having a direct transition between fermionic and trimer phases remains an open question in these models.

We acknowledge funding by LabEx PALM (ANR-10-LABX-0039-PALM). This work has been supported by Region Ile-de-France in the framework of the DIM Sirteq.

\bibliographystyle{apsrev4-2}
\bibliography{Trimers_v8.bib}

\onecolumngrid
\appendix
\newpage

\section*{Supplementary material for "Kinetic formation of trimers in a spinless fermionic chain"}
\renewcommand{\theequation}{S.\arabic{equation}}
 \setcounter{equation}{0}

\subsection{Effective free fermionic chain for $t=0$}

\begin{figure}[h]
\includegraphics[width=0.5\columnwidth]{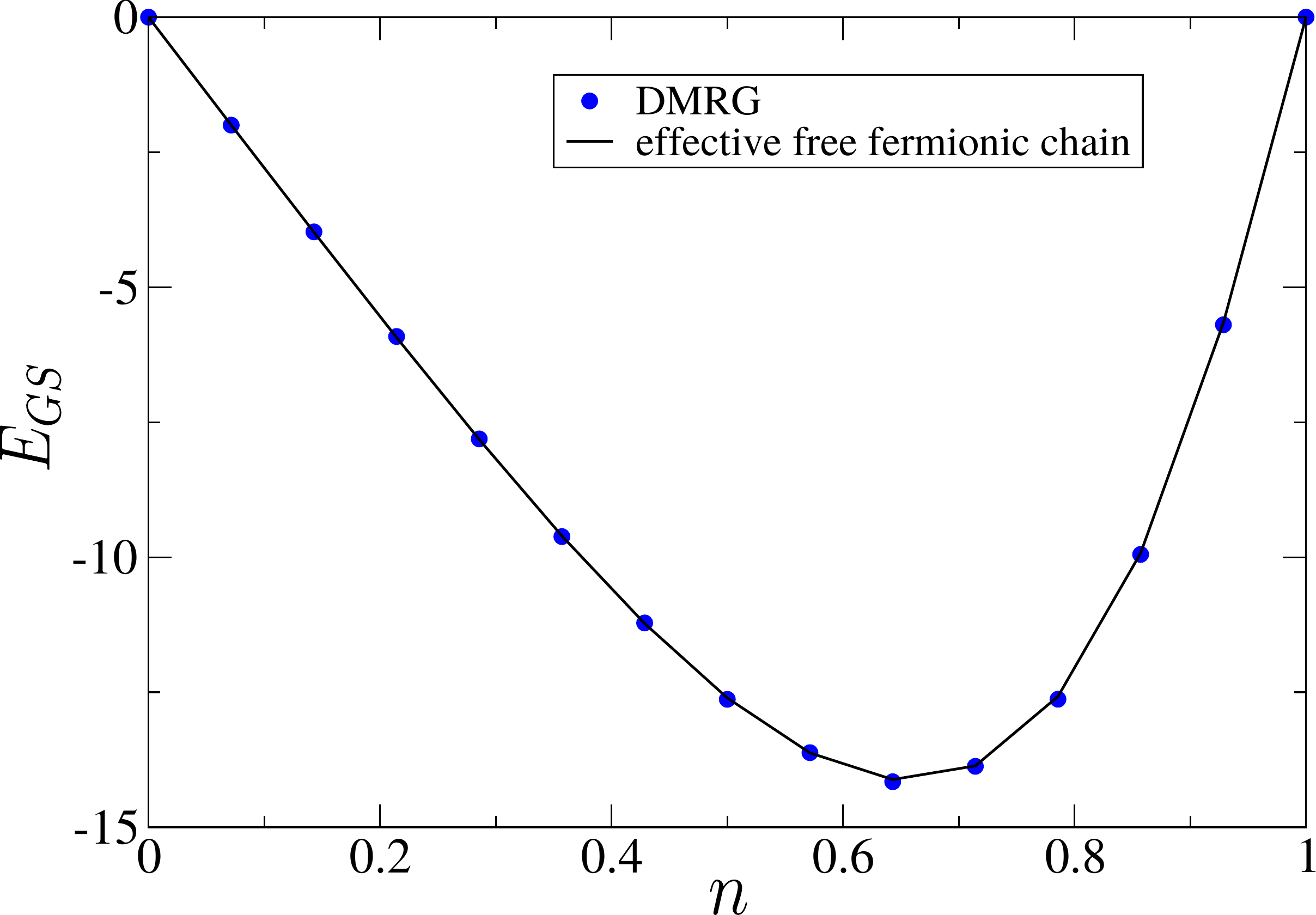}
 \caption{Ground state energy for $L=42$ sites in PBC with $t^{\prime}=1$ and $t=10^{-4}$ as a function of the filling compared with the analytical expression provided in Eq.~\eqref{Eq:gs_energy_T_phase}.}
 \label{Fig_SM:gs_energy_T_phase}
\end{figure}

After restricting the action of Hamiltonian~\eqref{Eq:Ham:GRA} of the main text to the subspace $\mathcal{H}_T$, the latter reduces to a free fermion model describing $N/3$ particles on a chain with $L-2N/3$ sites; the ground-state energy per site $\epsilon_{GS}=\langle\hat{H}\rangle/L$ when $t=0$ is given by:  
\begin{equation} \label{Eq:gs_energy_T_phase}
\epsilon_{GS}=-\frac{ 2|t^{\prime}|}{\pi}\left(1-\frac{2n}{3}\right)\sin\left( \frac{\pi n}{3-2n}\right).
\end{equation}
Equipped with Eq.~\eqref{Eq:gs_energy_T_phase} for the value of the ground state energy of the model in the low energy subspace $\mathcal H_T$, we compare it with its numerical estimate. We perform DMRG simulations on a system of size $L=42$ in PBC for $t^{\prime}=1$ and $t=10^{-4}$. A nonvanishing but negligible value of $t$ is used to allow the DMRG algorithm to converge to the ground state of the system irrespectively of the initial state. The result is presented in Fig.~\ref{Fig_SM:gs_energy_T_phase} as a function of the filling of the system and shows that the comparison is excellent.

\subsection{Generalized two-fluid model}

\begin{figure}[t]
\includegraphics[width=0.5\columnwidth]{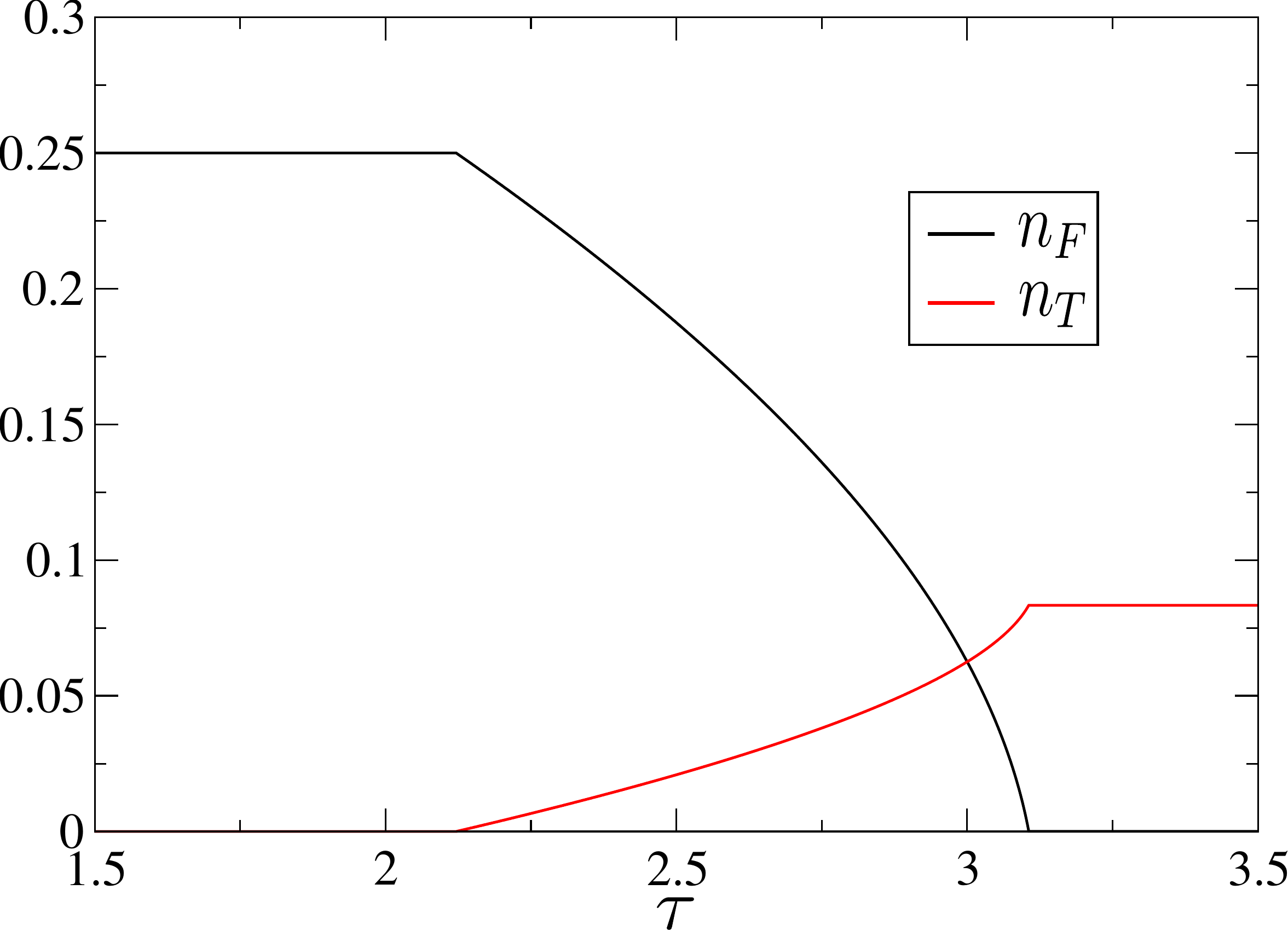}
\caption{Optimal fermionic density $n_F$ and trimer density $n_T=(n-n_F)/3$ as a function of $\tau$ obtained by minimizing the total energy $\langle\hat H_{2F,\,l=3}\rangle$ for a total density $n=0.25$. We observe (i) the $F$ phase ($n_F=n$) for $\tau<\tau_{c1}\approx 2.12$, (ii) the $T$ phase ($n_{F}=0$) for $\tau>\tau_{c2}\approx 3.11$ and (iii) the $TF$-$C$ phase ($0<n_{F}<n$) for $\tau_{c1}<\tau<\tau_{c2}$.}
\label{Fig_SM:2F_densities}
\end{figure}

We derive here the main predictions of model~\eqref{Eq:TwoFluids} of the main text for a coexistence phase between a liquid of uncoupled fermions and a liquid of trimers. For the sake of generality, we redefine model~\eqref{Eq:TwoFluids} of the main text in terms of a fluid of fermions and a fluid of $l$-mers (molecular fermions of size $l$). The explicit formulation of the model reads:
\begin{equation} \label{Eq_SM:2F_Hamiltonian}
 \hat H_{2F,\,l} = \sum_k \varepsilon_{F,k} \hat a_k^\dagger \hat a_k+\sum_k \varepsilon_{M,k}\hat d_k^\dagger \hat d_k,
\end{equation} 
where $\varepsilon_{M,k}=-2t^{\prime} \cos k$.
We minimize the total energy of Hamiltonian~\eqref{Eq_SM:2F_Hamiltonian} at fixed total filling $n_F+l n_M=n$. We are then able to obtain the population of the two species as a function of $\tau=|t^{\prime}/t|$.
We present in Fig.~\ref{Fig_SM:2F_densities} the general structure of the resulting phase diagram for the case $l=3$ by showing the fermionic density profile $n_F$ and the trimer density profile $n_T$ as a function of $\tau$: we observe the purely fermionic region ($n_{F}=n$) at small values of $\tau$ ($F$ phase), the purely trimer region ($n_{F}=0$) at large values of $\tau$ ($T$ phase) and the fermion-trimer coexistence region ($0<n_{F}<n$) for an intermediate range of values of $\tau$ ($TF$-$C$ phase) separating the two aforesaid phases. The three aforesaid phases generalize straightforwardly to any $l\geq 2$ case to (i) $F$ phase, where only the band $\varepsilon_{F,k}$ is populated, (ii) molecular ($M$) phase, where only $\varepsilon_{M,k}$ is populated, and (iii) molecule-fermion coexistence ($MF$-$C$) phase, where both bands are partially filled.

In order to determine the phase boundaries of the FM-C phase, it suffices to consider the stationarity condition determining the optimal fermionic density in the coexistence phase for the general case of molecules of size $l$, namely:
\begin{equation} \label{Eq_SM:stationarity_condition}
\cos(\pi n_{F})=\frac{\tau}{l}\cos\left(\pi\frac{n- n_{F}}{l}\right)
\end{equation}
The critical point $\tau_{c1}$ separating the $F$ phase from the $MF$-$C$ phase is obtained by imposing that Eq.~\eqref{Eq_SM:stationarity_condition} is satisfied for $\tau=\tau_{c1}$ when $n_f=n$. The resulting expression that one obtains is:
\begin{equation}
\tau_{c1}=l\cos(\pi n),
\end{equation} 
which in the case $l=3$, $n=1/4$ leads to the estimate $\tau_{c1}\approx 2.12$, whereas for $l=4$, $n=1/4$ it results in $\tau_{c1}\approx 2.83$.
Similarly, we find the critical point $\tau_{c2}$ separating the $MF$-$C$ phase from the $M$ phase by setting $\tau=\tau_{c2}$ and $n_f=0$ in Eq.~\eqref{Eq_SM:stationarity_condition}. The result is given by:
\begin{equation}
\tau_{c2}=\frac{l}{\cos\left(\frac{\pi n}{l}\right)},
\end{equation}
which in the case $l=3$, $n=1/4$ results in the estimate $\tau_{c2}\approx 3.11$, whereas for $l=4$, $n=1/4$ it gives $\tau_{c2}\approx 4.08$.

\subsection{Optimal fermion-trimer recombination process}

We now consider a Fermi sea of unpaired fermions with Fermi edges at $k=\pm \pi n$ and
study the formation of a trimer with momentum $k=0$ via a momentum-conserving process of annihilation of three fermions and creation of a trimer. We prove that the optimal way of creating a trimer with momentum $k=0$, i.e., at the bottom of the trimer band, via annihilation of $3$ fermions from a Fermi sea is realized when the the latter have momenta $k=0,\pm k_F$, where $k_F=\pi n$ is the Fermi momentum. To this goal, we need to determine a triplet of momenta $(k_1,k_2,k_3)\in [-\pi n, \pi n]^3$ satisfying $\sum_{i=1}^3 k_i=0$ such that the total energy loss $-2t\cos k_1-2t\cos k_2-2t\cos k_3$ due to the annihilation of fermions at momenta $k_1,k_2,k_3$ is maximal. By symmetry, we can always assume that $k_1\leq 0$ and $k_2,k_3\geq 0$. This simple observation allows the following manipulation:
\begin{subequations}
\begin{align}
&\max_{(k_1,k_2,k_3)\in [-\pi n, \pi n]^3:\sum_{i=1}^3 k_i=0}\left\{-2t\cos k_1-2t\cos k_2-2t\cos k_3\right\}=\\
&\max_{-\pi n\leq k_1\leq 0}\left\{-2t\cos k_1+\max_{0\leq k_2\leq -k_1}\{-2t\cos k_2-2t\cos (-k_1-k_2) \} \right\}.
\end{align}
\end{subequations} 
Since $\max_{0\leq k_2\leq -k_1}\{-2t\cos k_2-2t\cos (-k_1-k_2) \}$ is achieved by choosing $k_2=0$ (or equivalently $k_2=-k_1$), we are left with the problem of finding:
\begin{equation}
\max_{-\pi n\leq k_1\leq 0}\left\{-4t\cos k_1 \right\}-2t,
\end{equation} 
which manifestly leads to the optimal value $k_1=-\pi n$. The third momentum value is obtained from the constraint $k_3=-k_1-k_2=\pi n$ (or equivalently $k_3=0$ after the alternative choice $k_2=-k_1$), thus confirming the aforementioned optimal choice. 

This observation motivates the form of the BCS-like wavefunction proposed in Eq.~\eqref{Eq:BCS:3} of the main text. As the reader can observe, the creation of a trimer at momentum $k$ is accompanied in the variational ansatz by the annihilation of three fermions at momenta $k$ and $\sim \pm(k_F-2|k|)$, so that the identified most prominent correlation effects between the two fluids are qualitatively taken into account while preserving the analytical tractability of the trial wavefunction.

\subsection{The variational ansatz $| \Psi_3 \rangle$} 

The introduction of $\delta_k$ in the definition of the variational wavefunction $\ket{\psi_3}$ is justified by following the thinking process leading to Eq.~\eqref{Eq:BCS:3} of the main text. Indeed, we want to construct a state that gradually interpolates between a Fermi sea of fermions occupying momenta $-k_F<k<k_F$ and a Fermi sea of trimers occupying momenta $-\frac{k_F}{3}<k<\frac{k_F}{3}$ by means of the most relevant correlation effects between the two species. As the above calculation suggests, the latter are given by recombination processes where the creation of a trimer at momentum $k\in \left(-\frac{k_F}{3},\frac{k_F}{3}\right)$ is accompanied by the annihilation of fermions at momenta $-k_F+A|k|$, $k$, $k_F-A|k|$, $A$ being a proportionality constant. 

By imposing that the filling of the entire Fermi sea of trimers must correspond to the full depletion of the Fermi sea of fermions, one easily notices that it implies $A=2$. However, this condition alone is not sufficient for two reasons: (i) the fermionic states at momenta $\pm \left(k_F-2|k|+\frac{2\pi}{L}\right)$ (i.e., separated by an odd number of momentum quantization steps from $k_F$) are not getting depleted and (ii) the fermionic momenta getting depleted while filling trimer states at momenta $k$ and $-k$ coincide, thus hindering the analytical tractability of an ansatz in the form of $\ket{\psi_3}$. We solve these issues by depleting the momenta of the form $\pm \left(k_F-2|k|+\frac{2\pi}{L}\right)$ when filling the $k<0$ trimer states, relying on the fact that a shift of $\frac{2\pi}{L}$ in the chosen momenta will not impact the qualitative features of the results in the thermodynamic limit.

\subsection{Derivations within the $3BCS$ model}

\subsubsection{Expression for the energy density and optimal variational parameters}

As discussed in the main text, when the bands of fermions and trimers have both their minima at $k=0$, we need to enrich the noninteracting two-fluid model with an interspecies coupling term in order to take recombination processes into account.
The two-fluid model including an interaction term between fermions and trimers reads:

\begin{align}
\hat{H}=\sum_k(\epsilon_{k,f}-\mu)\hat a^{\dag}_{k} \hat a_{k}+\sum_k(\epsilon_{k,t}-3\mu)\hat d^{\dag}_{k}\hat d_{k}+\frac{ig}{3L} \sum_{k_1,k_2,k_3} f(k_1,k_2,k_3) \hat d_{k_1+k_2+k_3}^\dagger 
 \hat a_{ k_1} 
 \hat a_{ k_2} 
\hat a_{ k_3} + H.c.,
\end{align}
where the last term is the reciprocal space representation of $\hat H_{int}=g\sum_j \hat d^{\dag}_j \hat a_{j-1} \hat a_j \hat a_{j+1}+H.c.$ and $f(k_1,k_2,k_3)=\sin(k_3-k_1)+\sin(k_2-k_3)+\sin(k_1-k_2)$. 

The evaluation of the expectation value of the energy over the variational state $\ket{\Psi_3}$ gives as a result:
\begin{align}
&\expval{H}_{\Psi_3}-E_{FS}=\\
&\sum_{0\leq k<\frac{k_F}{3}}\left[(\epsilon_{k,t}-\epsilon_{k,F}-\epsilon_{k_F-2k,F}-\epsilon_{-k_F+2k,F})|\beta_k|^2+\frac{4g}{L}f(k,k_F-2k,-k_F+2k)Im\{\alpha_k^{*}\beta_k\} \right]+\nonumber\\
&\sum_{-\frac{k_F}{3}< k<0}\Biggl[(\epsilon_{k,t}-\epsilon_{k,F}-\epsilon_{k_F+2k+\frac{2\pi}{L},F}-\epsilon_{-k_F-2k-\frac{2\pi}{L},F})|\beta_k|^2+\frac{4g}{L}f\left(k,k_F+2k+\frac{2\pi}{L},-k_F-2k-\frac{2\pi}{L}\right)Im\{\alpha_k^{*}\beta_k\} \Biggr],\nonumber
\end{align}
where $E_{FS}$ is the energy of the fermionic Fermi sea filled up to the Fermi momentum $k_F$. 
At this point, we perform the following approximations: (i) we plug in the expressions $\epsilon_{k,F}=-2t\cos{k}$ and $\epsilon_{k,t}=-2t^{\prime}\cos{k}$ for the noninteracting dispersion relations of fermions and trimers; (ii) due to the normalization condition, we parametrize the unknown coefficients as $\alpha_k=\cos{\theta_k}$, $\beta_k=e^{i\varphi_{k}}\sin{\theta_k}$, as detailed in the main text; (iii) assuming that the system size $L$ is sufficiently large, we perform the replacement $\sum_k\rightarrow \frac{L}{2\pi}\int dk$, we neglect the shifts by $\frac{2\pi}{L}$ and we change variable according to $k\rightarrow -k$ in the resulting integral ranging over negative values of $k$; as a result one obtains Eq.~\eqref{Eq:Energy:Functional} of the main text.

The expression in Eq.~\eqref{Eq:Energy:Functional} of the main text is a functional of the variational parameters $\theta_k$ and $\phi_k$, but it is trivial, as it does not feature terms coupling the unknown functions $\theta_k$ and $\phi_k$ in a nonlocal way. Thus, its minimization boils down to the minimization of the integrand function with respect to the two variables $\theta_k$ and $\phi_k$ for each value of $k$ separately. A standard calculation leads to the optimal values displayed in Eq.~\eqref{Eq:Optimal_Params} of the main text.

\begin{figure}[t]
\includegraphics[width=0.5\columnwidth]{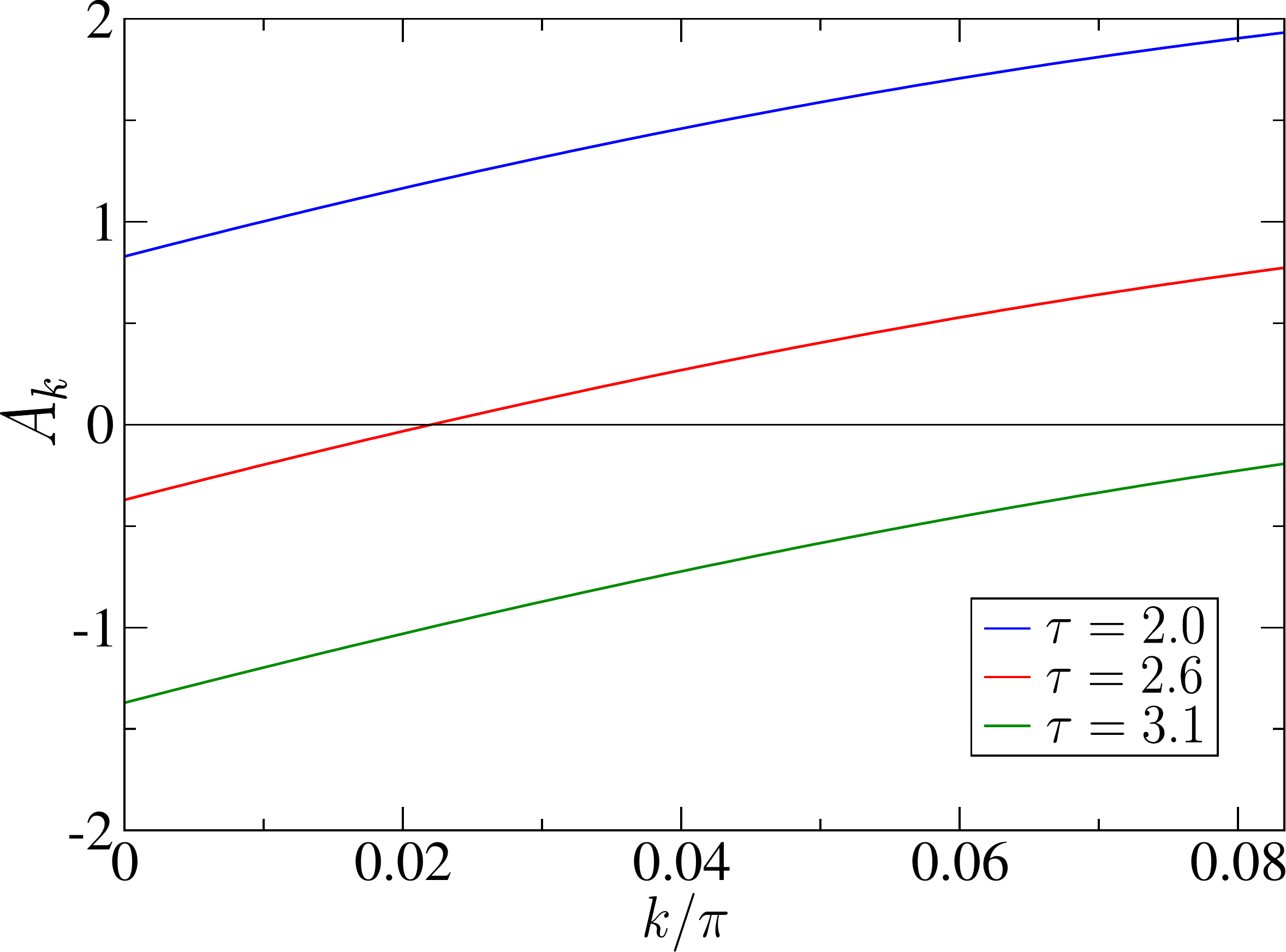}
\caption{$A_k$ as a function of $k$ for $\tau<\tau_1$ (blue curve), $\tau_1<\tau<\tau_2$ (red line) and $\tau>\tau_2$ (green line).}
\label{Fig_SM:A(k)}
\end{figure}

The value of the optimal parameters depends crucially on the sign of $A_k$. Therefore, we plot in Fig.~\ref{Fig_SM:A(k)} the function $A_k$ in the interval $0\leq k\leq k_F/3$ in three cases, representing respectively the small $\tau$, intermediate $\tau$ and large $\tau$ behavior of $A_k$. We observe that a small value of $\tau$, where we expect to predict the $F$ phase, is linked to a strictly positive $A_k$ profile, while a large value of $\tau$, where the $T_0$ phase is supposed to appear, is associated to a strictly negative $A_k$ profile. 

In an intermediate regime of $\tau$ values, instead, $A_k$ changes sign at a value of $k\in [0,k_F/3]$ that continuously grows from $k=0$ to $k=k_F/3$. The boundaries of the interval of $\tau$ values where this happens are found by determining the values $\tau_1$ and $\tau_2$ of $\tau$ such that $A_{k=0}$ and $A_{k=k_F/3}$ vanish, respectively. The result reads:
\begin{align} 
&\tau_{1}=1+2\cos(\pi n),\\
&\tau_{2}=3.\nonumber
\end{align}

\subsubsection{Behavior of the trimer density}

In order to properly identify the phases predicted by the variational ansatz, we compute the trimer density as a function of $\tau$ in the thermodynamic limit. The expression of this observable is easily written as:
\begin{equation}
n_T=\frac{1}{L}\sum_k\langle \hat d^{\dag}_{k}\hat d_{k}\rangle=\frac{1}{\pi}\int_{0}^{\frac{k_F}{3}}dk \sin^{2}\theta_k.
\end{equation}

Since in the thermodynamic limit $\theta_k=\frac{\pi}{2}\Theta(-A(k))$ because $B_k\propto L^{-1}$, the evaluation of $n_T$ becomes trivial.
In particular, when $\tau<\tau_1$ one has a strictly positive $A(k)$ profile, which implies $\theta_k=0$ and thus $n_T=0$. Thus, we identify this regime with the $F$ phase. In a similar manner, since $A(k)$ is strictly negative when $\tau>\tau_2$, we get $\theta_k=\frac{\pi}{2}$, which results in $n_T=n/3$. We link this result with the $T_0$ phase. Finally, when $\tau_1<\tau<\tau_2$, one has $A(k)<0$ for $0<k<K(\tau)$ and $A(k)>0$ for $K(\tau)<k<k_F/3$, where $K(\tau)$ denotes the intermediate zero of $A(k)$ as a function of $\tau$. In this case, the trimer density takes the form:
\begin{equation}
n_T=\frac{1}{\pi}\left(\int_{0}^{K(\tau)}dk \sin^{2}\theta_k+\int_{K(\tau)}^{\frac{k_F}{3}}dk \sin^{2}\theta_k\right)=\frac{1}{\pi}\int_{0}^{K(\tau)}dk=\frac{K(\tau)}{\pi}.
\label{Eq:K(tau)}
\end{equation}
Hence, the trimer density takes an intermediate value $0<n_T<n/3$, which implies that the fermionic density $n_F=n-3n_T$ is nonvanishing as well and the two species coexist, as expected for a $TF$-$H$ phase. Eq.~\eqref{Eq:K(tau)} forces the interpretation of the zero of $A(k)$ as the Fermi momentum of the trimer Fermi sea as a function of $\tau$.

\subsubsection{Critical behavior of the energy density}

The behavior of the energy density in the thermodynamic limit can be as well evaluated in the $3$ phases. We assume in the following that the $g$ does not scale extensively as the system size.
In the $F$ phase, $A_k>0\,\,\,\forall k\in [0,\pi n/3]$, implying that $\theta_k=\frac{1}{2}\arctan\left(\frac{2B_k}{A_k}\right)$ and therefore resulting in the optimal energy:
\begin{align}
\frac{\langle{\hat H}\rangle_{\Psi_3}-E_{FS}}{Lt}=2\int_0^{\frac{k_F}{3}}\frac{dk}{2\pi}\Biggl\{A_k\sin^2 \left[\frac{1}{2}\arctan\left(\frac{2B_k}{A_k}\right)\right]-B_k\sin\left[\arctan\left(\frac{2B_k}{A_k}\right)\right]\Biggr\}.
\end{align} 
As $L\rightarrow +\infty$, $B_k\propto L^{-1}$ vanishes and therefore:
\begin{equation}
\lim_{L\to +\infty}\frac{\langle{\hat H}\rangle_{\Psi_3}-E_{FS}}{Lt}=0,
\end{equation}
indicating that the energy of the system equals the energy of the fermionic Fermi sea.

In the $T_0$ phase, instead, one has $A_k<0\,\,\,\forall k\in [0,\pi n/3]$, which implies that $\theta_k=\frac{1}{2}\arctan\left(\frac{2B_k}{A_k}\right)+\frac{\pi}{2}$ and gives as a result the optimal energy:
\begin{align}
\frac{\langle{\hat H}\rangle_{\Psi_3}-E_{FS}}{Lt}=2\int_0^{\frac{k_F}{3}}\frac{dk}{2\pi}\Biggl\{A_k\sin^2 \left[\frac{\pi}{2}+\frac{1}{2}\arctan\left(\frac{2B_k}{A_k}\right)\right]-B_k\sin\left[\pi+\arctan\left(\frac{2B_k}{A_k}\right)\right]\Biggr\}.
\end{align} 
As $L\rightarrow +\infty$, the second term clearly vanishes, as it is bounded from above by an expression proportional to $L^{-1}$, while the first term converges to $A_k$ and gives:
\begin{align}
\lim_{L\to +\infty}\frac{\langle{\hat H}\rangle_{\Psi_3}-E_{FS}}{Lt}=\frac{1}{\pi}\int_0^{\frac{k_F}{3}}A_k=-\frac{2}{\pi}\left(\tau-\frac{\sin(\pi n)}{\sin\left(\frac{\pi n}{3}\right)}\right)\sin\left(\frac{\pi n}{3}\right)
\end{align}

In the intermediate regime $\tau_1<\tau<\tau_2$, one has that $A_k<0\,\,\,\forall k\in [0,K(\tau))$ and $A_k>0\,\,\,\forall k\in (K(\tau),\pi n/3]$; therefore, the optimal energy takes the form:
\begin{align}
&\frac{\langle{\hat H}\rangle_{\Psi_3}-E_{FS}}{Lt}=2\int_0^{K(\tau)}\frac{dk}{2\pi}\Biggl\{A_k\sin^2 \left[\frac{\pi}{2}+\frac{1}{2}\arctan\left(\frac{2B_k}{A_k}\right)\right]-B_k\sin\left[\pi+\arctan\left(\frac{2B_k}{A_k}\right)\right]\Biggr\}+\\
&+2\int_{K(\tau)}^{\frac{k_F}{3}}\frac{dk}{2\pi}\Biggl\{A_k\sin^2 \left[\frac{1}{2}\arctan\left(\frac{2B_k}{A_k}\right)\right]-B_k\sin\left[\arctan\left(\frac{2B_k}{A_k}\right)\right]\Biggr\}.
\end{align}
The second term goes to zero in the large size limit, while the first term gives:
\begin{align}
\lim_{L\to +\infty}\frac{\langle{\hat H}\rangle_{\Psi_3}-E_{FS}}{Lt}=\frac{1}{\pi}\int_0^{K(\tau)}A_k=\frac{2}{\pi}(1-\tau)\sin{K(\tau)}+\frac{2}{\pi}\sin(\pi n)-\frac{2}{\pi}\sin\left[\pi n-2K(\tau)\right].
\end{align}
The asymptotic behavior of the energy as it approaches the critical points can then be obtained by deriving the one of the zero $K(\tau)$ when $\tau$ is close to either of the critical points; the latter is found by expanding the condition $A(k)=0$ around $k=0$ and $k=\frac{\pi n}{3}$, obtaining:
\begin{align}
&K(\tau)\sim\frac{\tau-\tau_{c1}}{4\sin(\pi n)}\\
&K(\tau)\sim\frac{\pi n}{3}-\frac{1}{6}\cot\left(\frac{\pi n}{3}\right)(\tau_{c2}-\tau)
\end{align}
As a result, in the limit $\tau\rightarrow\tau_{1}^+$, the energy density behaves as:
\begin{equation}
\frac{\langle{\hat H}\rangle_{\Psi_3}-E_{FS}}{Lt}\approx-\frac{(\tau-\tau_{c1})^2}{4\pi\sin(\pi n)},
\end{equation}
while in the limit $\tau\rightarrow\tau_{2}^-$, the energy density behaves as:
\begin{equation}
\frac{\langle{\hat H}\rangle_{\Psi_3}-E_{FS}}{Lt}\approx\frac{2}{\pi}\sin(\pi n)-\frac{6}{\pi}\sin\left(\frac{\pi n}{3}\right)+\frac{2}{\pi}\sin\left(\frac{\pi n}{3}\right)(\tau_{c2}-\tau)-\frac{1}{6\pi}\frac{\cos^2\left(\frac{\pi n}{3}\right)}{\sin\left(\frac{\pi n}{3}\right)}(\tau_{c2}-\tau)^{2}.
\end{equation} 
Hence, we recover the finite jump discontinuities in the second derivative associated to the appearance/disappearance of a gapless mode when entering/exiting a coexistence phase of fermions and trimers, as expected from the vanishing of the interspecies interaction energy in the thermodynamic limit.

We remark that the assumption that $g=O(1)$ is crucial in deriving the results above. As shown in Fig.~\ref{Fig:BCS:Study} of the main text, the predictions of the $3BCS$ model replicate in a reliable manner the features of the DMRG data as for qualitative aspects of the energy first derivative and its scaling with $L$. Nevertheless, the numerical data do not show undisputedly the critical behavior predicted by the $3BCS$ model and thus we are unable to rule out with absolute certainty a different kind of criticality separating the $F$ phase from the $TF$-$H$ phase.

\begin{figure}[h]
\includegraphics[width=0.9\columnwidth]{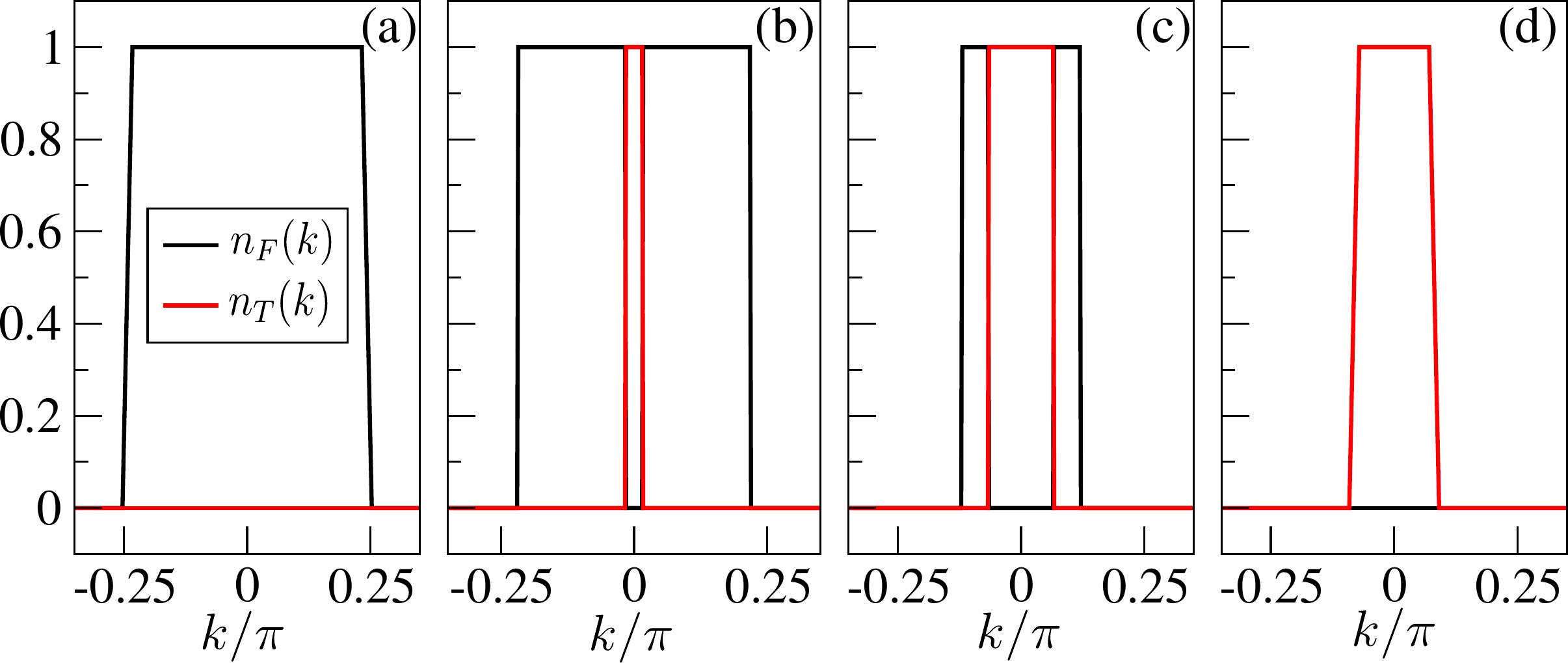}
 \caption{Occupation functions $n_F(k)$ and $n_T(k)$ in the thermodynamic limit in the (a) $F$ phase for $\tau=2.0$, (b) $TF$-$C$ phase for $\tau=2.55$, (c) $TF$-$C$ phase for $\tau=2.9$ and (d) $T_0$ phase for $\tau=3.1$.}
 \label{Fig:_SM:occ_functions_3BCS}
\end{figure}

\subsubsection{Occupation functions}

We characterize here the behavior of the fermionic occupation function $n_F(k)=\langle \hat a^{\dag}_k \hat a_k\rangle$ and of the trimer occupation function $n_T(k)=\langle \hat d^{\dag}_k \hat d_k\rangle$. The fermionic occupation function takes the form:

\begin{equation}
n_F(k)=\begin{cases}
\cos^2\theta_{|k|} & |k|<\frac{k_F}{3}\\
\cos^2\theta_{\frac{k_F-|k|}{2}} & \frac{k_F}{3}<|k|<k_F,
\end{cases}
\end{equation}
whereas the trimer occupation function reads:
\begin{equation}
n_T(k)=\sin^2\theta_{|k|},\,\,\, |k|<\frac{k_F}{3}.
\end{equation}
Thus, in the thermodynamic limit we can identify the functions $n_F (k)$ and $n_T(k)$ in the $3$ phases of the model, as shown in Fig.~\ref{Fig:_SM:occ_functions_3BCS}(a-d): in the $F$ phase, $\theta_k=0$ for every value of $k$ and $n_F(k)=\mathbf{1}_{[-k_F,k_F]}(k)$, recovering the standard Fermi sea filled up to momentum $k_F$, while $n_T(k)$ vanishes; in the $TF$-$H$ phase, $\theta_k=\frac{\pi}{2}$ for $0<k<K(\tau)$ and $\theta_k=0$ for $K(\tau)<k<\frac{k_F}{3}$, and therefore one gets $n_F(k)=\mathbf{1}_{[-k_F+2K(\tau),K(\tau)]\cup[K(\tau),k_F-2K(\tau)]}(k)$ and $n_T(k)=\mathbf{1}_{[-K(\tau),K(\tau)]}(k)$ (recovering the qualitative structure of the numerical data in Fig.~\ref{Fig:nk} of the main text); in the $T_0$ phase, instead, $\theta_k=\frac{\pi}{2}$ for every value of $k$, resulting in a vanishing $n_F(k)$ and $n_T(k)=\mathbf{1}_{[-k_F/3,k_F/3]}(k)$, i.e., a Fermi sea of trimers at density $\frac{n}{3}$.
   
\subsection{Finite size effects from the variational energy density}

The variational expression of Eq.~\eqref{Eq:Energy:Functional} of the main text for the energy density holds in the limit of large sizes, as it was constructed by replacing discrete sums over momenta with continuous integrals via the rule $\sum_k\rightarrow\frac{L}{2\pi}\int dk$. 
If we take the thermodynamic limit of the 3BCS model,
the ratio $g/L$ tends to zero
and the minimization of the energy functional becomes particularly simple:
$\varphi_k$ is unconstrained and $\theta_k $ takes only two values: $0$ when $A_k>0$ and $\pi/2$ when $A_k<0$. Thus, for each value of $k$, either $\alpha_k=1$, $\beta_k=0$ or vice versa, and  the ansatz becomes a product state in momentum space, i.e., an uncorrelated state of delocalized fermions and trimers. 

\begin{figure}[h]
\includegraphics[width=0.8\columnwidth]{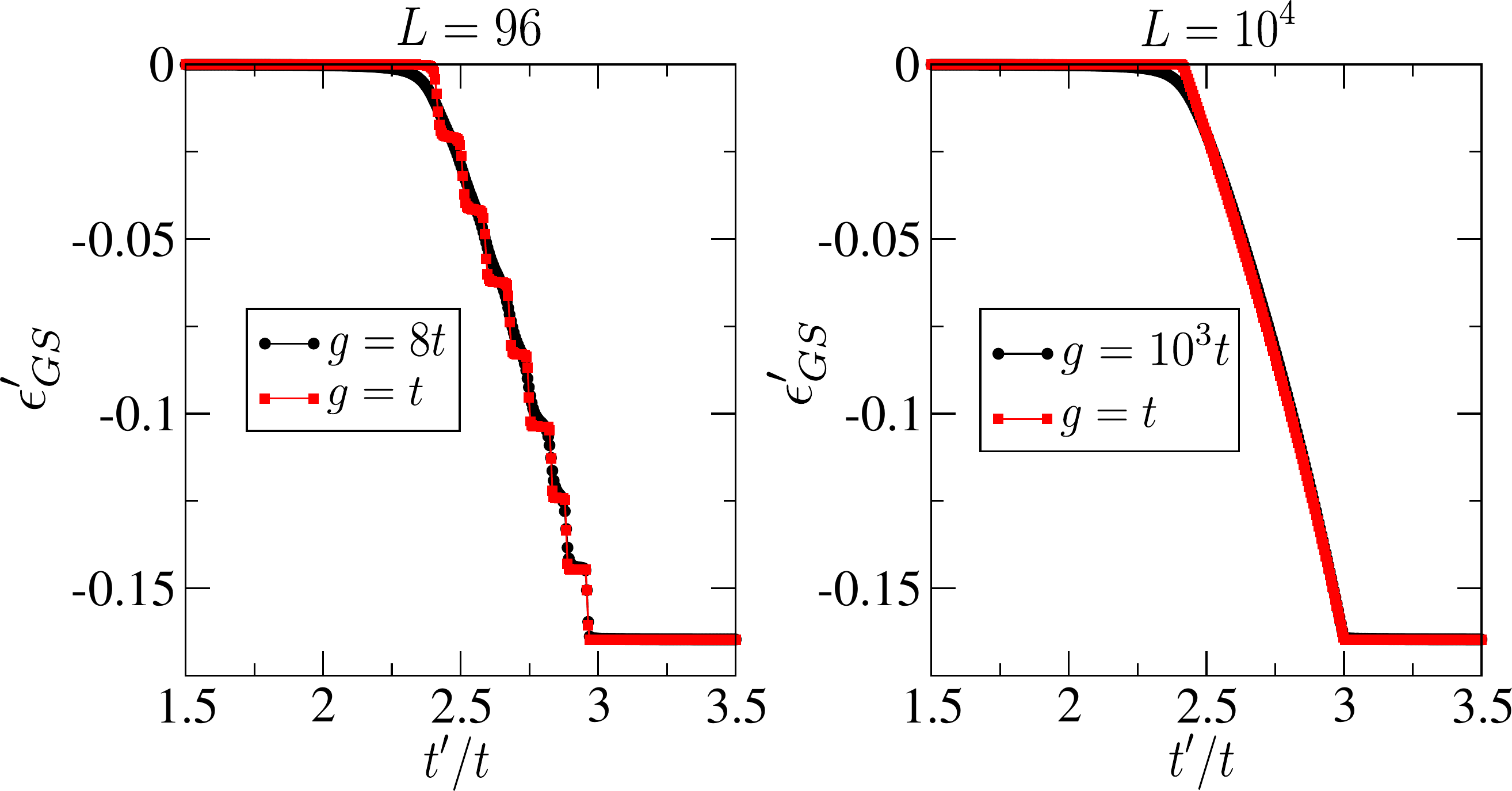}
 \caption{(a) First derivative of the energy density within our BCS-like approach for $g=t,\, 8t$ and $L=96$. (b) First derivative of the energy density within our BCS-like approach for $g=t,\, 10^3 t$ and $L=10^4$.}
 \label{Fig_SM:finite_size}
\end{figure}

Therefore, we need to clearly identify which finite size effects are induced by the presence of a nonvanishing value of $\frac{g}{Lt}$ and which ones appear as a result of a small value of $L$. The answer to this question is presented in Fig.~\ref{Fig_SM:finite_size}(a-b), where the profile of the first derivative of the variational ground state energy density is shown for different choices of $L$ and $g$. When the value of $L$ is within reach of numerical simulations but not necessarily sufficient to display clear thermodynamic limit behavior (panel (a)), the comparison between the curves corresponding to different values of hybridization proves that trimer-fermion coupling smoothens out the profile around the critical point separating the F phase from the FT-H phase; on the other hand, the step-like behavior of the curves for $L=96$ is present irrespectively of the value of $g$ and is purely a finite-size effect, as it disappears from the plots displayed in panel (b) for $L=10^4$. Furthermore, the effect of a sufficiently large value of $g$ relative to $Lt$ replicates the phenomenology observed at finite size, thus confirming the aforementioned conclusion on the effect of the interspecies coupling $g$.
\subsection{Occupation factors}
\begin{figure}[h]
\includegraphics[width=0.8\columnwidth]{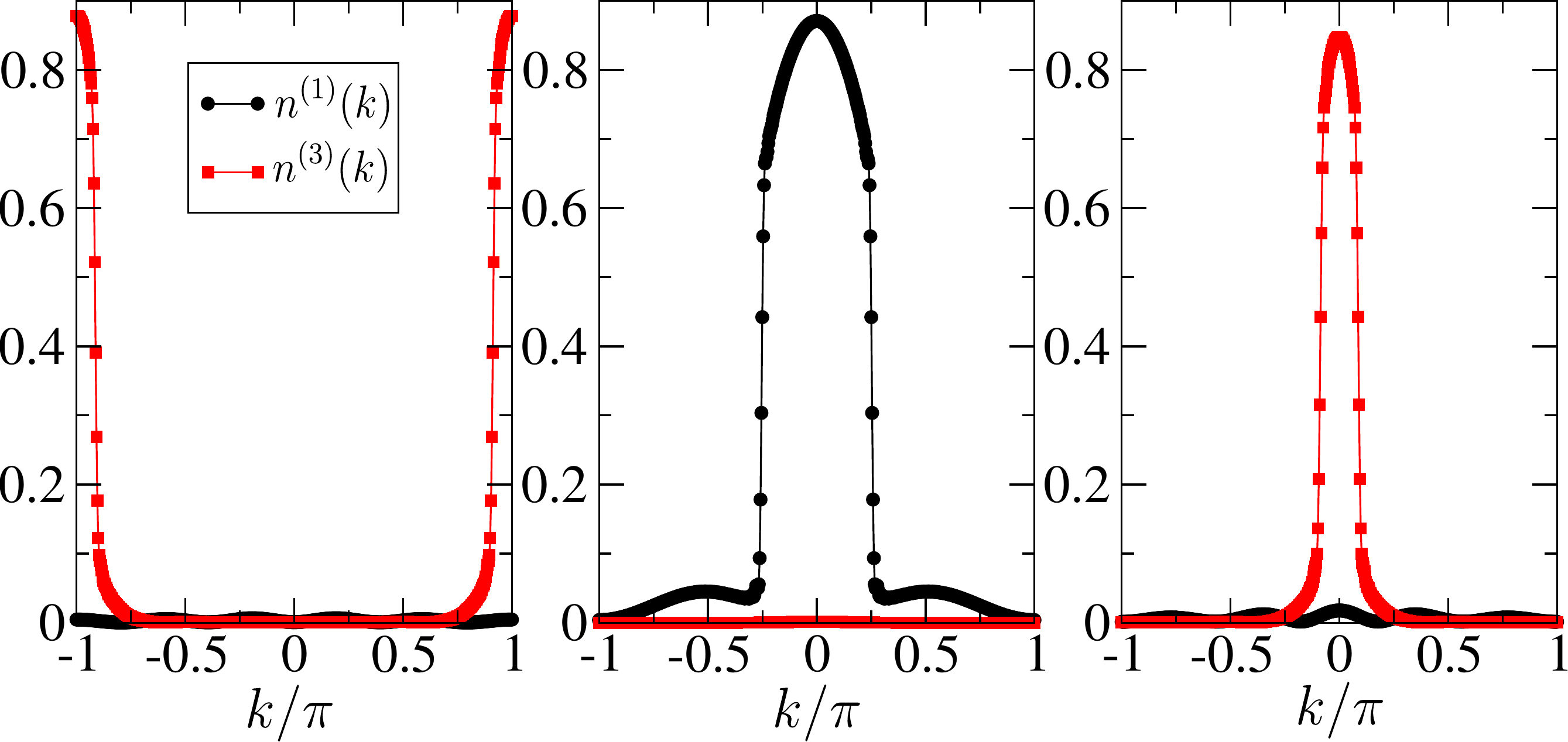}
 \caption{(a) Momentum distribution functions $n^{(1)}(k)$ and $n^{(3)}(k)$ for $t^{\prime}/t=-3.5$ (T$_\pi$ phase) with $L=96$ in OBC. (b) Momentum distribution functions $n^{(1)}(k)$ and $n^{(3)}(k)$ for $t^{\prime}/t=1.0$ (F phase) with $L=96$ in OBC. (c) Momentum distribution functions $n^{(1)}(k)$ and $n^{(3)}(k)$ for $t^{\prime}/t=3.5$ (T$_0$ phase) with $L=96$ in OBC.}
 \label{Fig_SM:occupation_factors}
\end{figure}

The nonstandard operators $\hat F_j^{(M)}$ defined in Eq.~\eqref{Eq:Projected_Operators} of the main text allow to target the momentum occupation function $n^{(M)}(k) = \sum_{j,l} e^{- i k (j-l)} \langle \hat F^{(M)\dagger}_j \hat F^{(M)}_l\rangle$ of unbound fermions and trimers, corresponding respectively to the choices $M=1$ and $M=3$. 
We complement the information contained in Fig.~\ref{Fig:nk} of the main text by providing in Fig.~\ref{Fig_SM:occupation_factors}(a-c) the uncoupled fermion and trimer momentum occupation functions, $n^{(1)}(k)$ and $n^{(3)}(k)$ respectively, for the T$_\pi$, T$_0$ and F phases. The T$_\pi$ and T$_0$ phases feature a negligible $n^{(1)}(k)$ profile together with a quasicondensate around $k=\pi$ and $k=0$, respectively, in the trimer occupation function profile $n^{(3)}(k)$. On the other hand, the F phase shows a standard Fermi sea of uncoupled fermions around $k=0$ in the uncoupled fermion occupation function $n^{(1)}(k)$ and a vanishing $n^{(3)}(k)$ profile. These results are in striking agreement with the effective two-fluid approach adopted in the analysis of the phase diagram, as they give direct evidence for the occupation of the effective bands for uncoupled fermions and trimers in the single species phases to be found while varying the ratio $t^{\prime}/t$.

\subsection{Fourier transform of the density profile}

\begin{figure}[h]
\includegraphics[width=0.9\columnwidth]{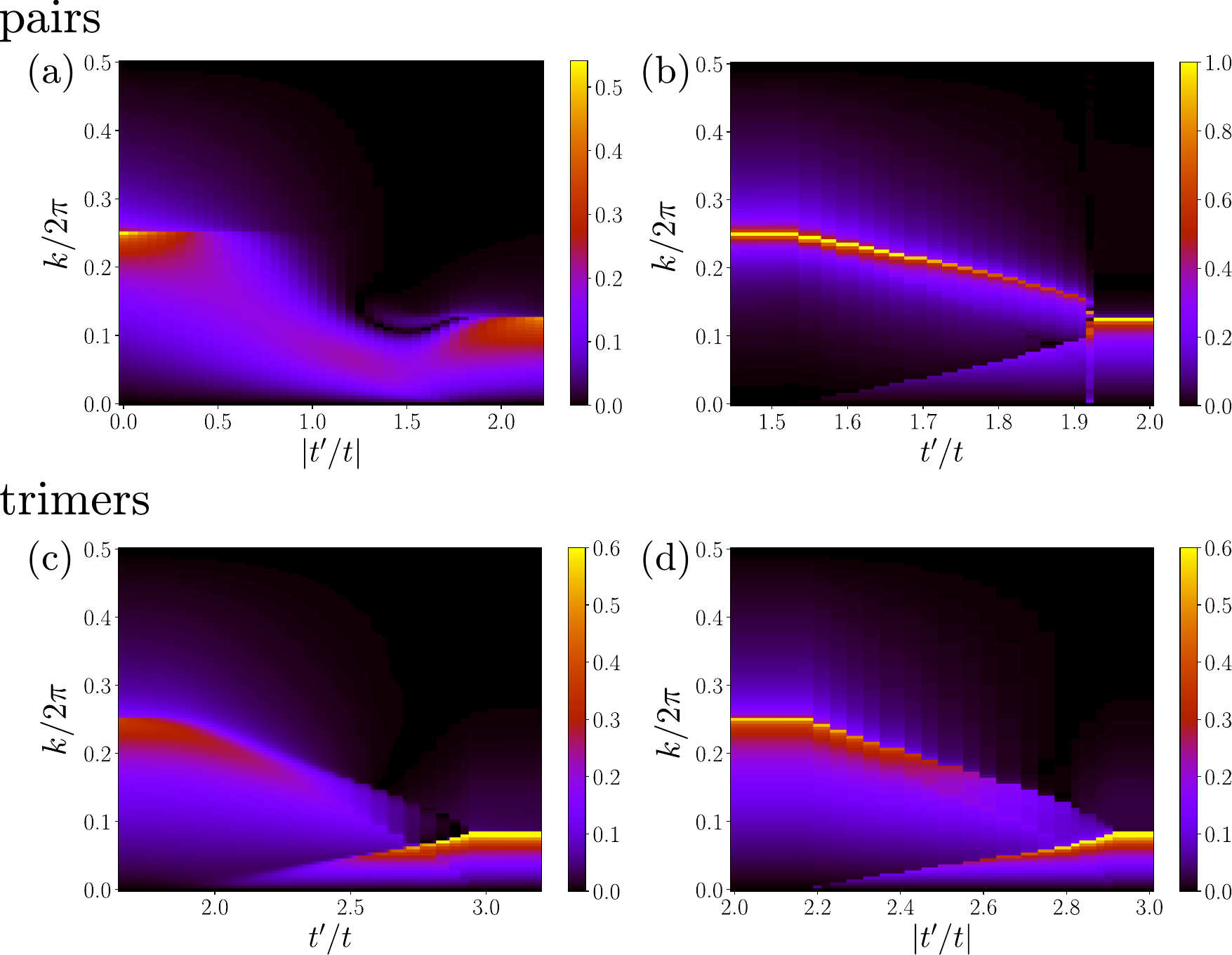}
 \caption{(a-b) Fourier transform of the density profile in the ground state of the model with correlated pair hopping studied in~\cite{Gotta_2021,Gotta_2021_B} for $t^{\prime}/t<0$ (a) and $t^{\prime}/t>0$ (b). (c-d) Fourier transform of the density profile in the ground state of the model defined in Eq.~\eqref{Eq:Ham:GRA} of the main text  for $t^{\prime}/t>0$ (c) and $t^{\prime}/t<0$ (d).}
 \label{Fig_SM:N(k)}
\end{figure}

We analyze the behavior of the Fourier transform of the density profile in the parameter regions where the transition from T$_\pi$ and T$_0$ phases to the F phase takes place, respectively. The result, presented in Fig.~\ref{Fig_SM:N(k)}(c-d), shows that the behavior in the two cases is qualitatively similar.

In the fermionic phase, a peak at $k=2\pi n$ is observed, in agreement with the Luttinger liquid prediction of the wavevector associated with the leading modulation in the density fluctuations. Similarly, in the trimer phase the effective Luttinger liquid density is $\frac{n}{3}$ due to trimer formation, thus resulting in a leading peak at $k=2\pi \frac{n}{3}$.

The situation is richer in the intermediate phase separating the two aforesaid limits. We first observe a peak interpolating between the one in the F phase and the one in the T$_\pi$ and T$_0$ phases. If we call $n_F$ and $n_T$ the effective densities of uncoupled fermions and trimers in the system, respectively, this signal can be interpreted as the leading density modulation in a Luttinger liquid phase at density $n_F+n_T$ where both uncoupled fermions and trimers populate the lattice. The peak is less sharply defined close to the transition from the F phase to the FT-H phase, where the two species are strongly hybridized. Consistently with the picture of a coexistence between uncoupled fermions and trimers, we additionally observe a peak due to trimer formation at $k=2\pi n_T$, which vanishes in the F phase and connects to the value $k=2\pi\frac{n}{3}$ when approaching the T$_\pi$ and T$_0$ phases, and a subleading peak at $k=2\pi n_F$ which goes to zero when entering the T$_\pi$ and T$_0$ phases. 

For completeness, we provide in Fig.~\ref{Fig_SM:N(k)}(a-b) the same quantity for the ground state of the Hamiltonian studied in~\cite{Gotta_2021, Gotta_2021_B}, whose interaction term implements correlated pair hopping processes. For $t^{\prime}/t>0$, a coexistence phase between paired and unpaired fermions is observed. If we denote the effective pair density as $n_P$, it results in the appearance of two distinct peaks at $k=2\pi (n_F+n_P)$ and $k=2\pi n_P$, together with a subleading one at $k=2\pi n_F$. On the other hand, when $t^{\prime}/t<0$, a critical point belonging to the Ising universality class separates the $F$ phase from a Luttinger liquid of pairs. The latter originates as a result of relevant pair-fermion interactions in an effective two-fluid description. Thus, it is not possible to observe an extended parameter region featuring well-defined peaks associated to effective gapless modes for pairs and unpaired fermions, respectively.

\end{document}